\documentclass[journal]{vgtc}                     

\onlineid{0}

\vgtccategory{Research}

\vgtcpapertype{theory/model}

\title{Modeling the Dashboard Provenance}

\author{%
  \authororcid{Johne Jarske}{0000-0001-8907-6455}, and
  \authororcid{Jorge Rady}{0000-0003-3839-4570}, and 
  \authororcid{Lucia V. L.\ Filgueiras}{0000-0003-3791-6269}, and \\
  \authororcid{Leandro M.\ Velloso}{0000-0003-4883-7208}, and 
  \authororcid{Tania L.\ Santos}{0000-0001-6912-6793}  
}

\authorfooter{
  \item
  	Johne M.\ Jarske, Jorge Rady, Lucia V. L.\ Filgueiras and Leandro M.\ Velloso are with Universidade de São Paulo.
  	email: \{johne.jarske\,$|$\, jorgerady\,$|$\, lfilguei\,$|$\, leandrovelloso\}@usp.br\,.
  \item	
  	Tania L.\ Santos is with Faculdade de Tecnologia de Cotia.
	email: tania.ls@outlook.com.br
}

\abstract{
Organizations of all kinds, whether public or private, profit-driven or non-profit, and across various industries and sectors, rely on dashboards for effective data visualization. However, the reliability and efficacy of these dashboards rely on the quality of the visual and data they present. Studies show that less than a quarter of dashboards provide information about their sources, which is just one of the expected metadata when provenance is seriously considered. Provenance is a record that describes people, organizations, entities, and activities that had a role in the production, influence, or delivery of a piece of data or an object. This paper aims to provide a provenance representation model, that entitles standardization, modeling, generation, capture, and visualization, specifically designed for dashboards and its visual and data components. The proposed model will offer a comprehensive set of essential provenance metadata that enables users to evaluate the quality, consistency, and reliability of the information presented on dashboards. This will allow a clear and precise understanding of the context in which a specific dashboard was developed, ultimately leading to better decision-making.  
}


\teaser{
  \centering
  \includegraphics[width=\linewidth]{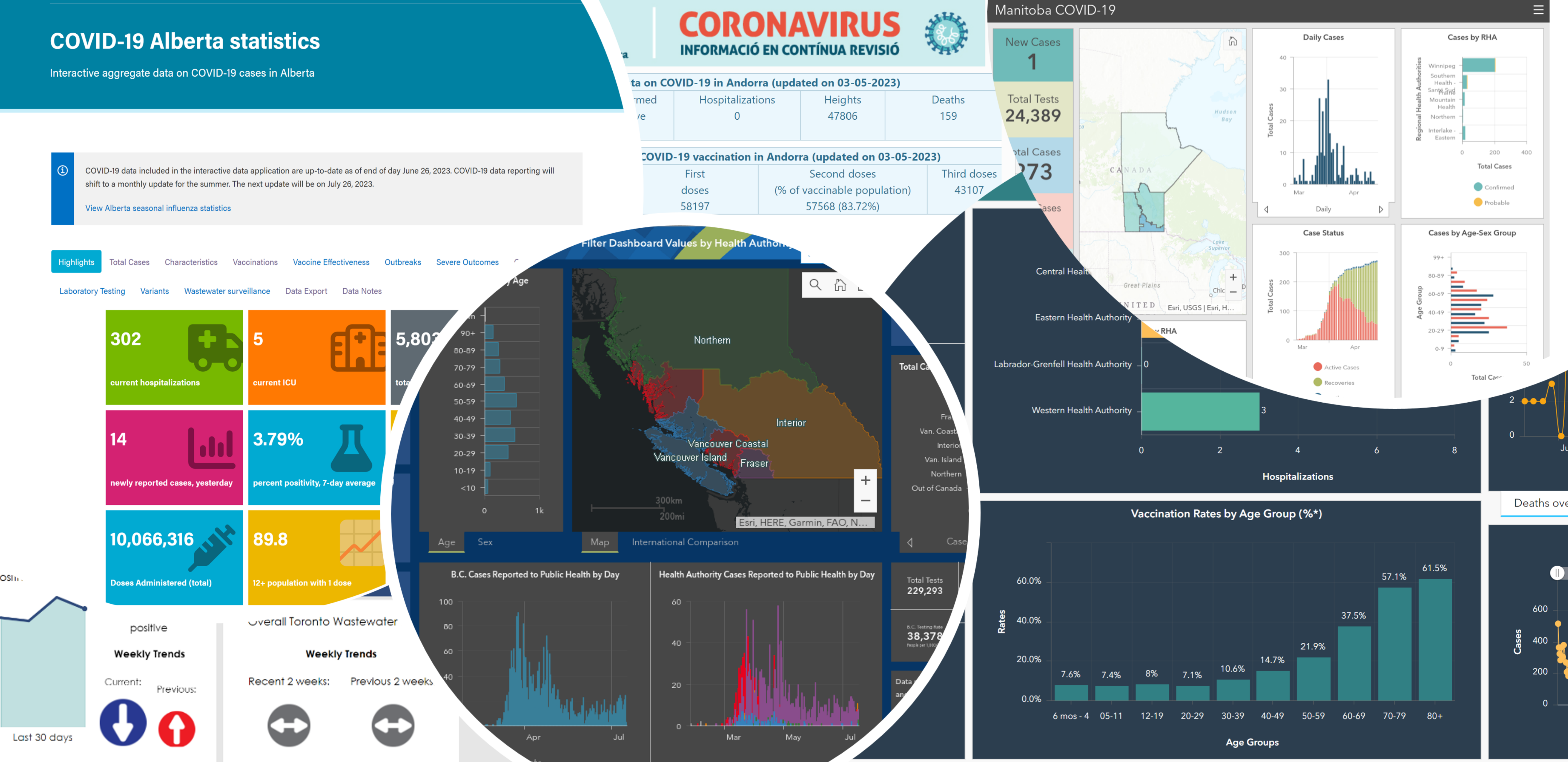}
  \caption{%
  	Web-based COVID-19 Dashboards.%
  }
  \label{fig:teaser}
}

\graphicspath{{figs/}{figures/}{pictures/}{images/}{./}} 

\usepackage{tabu}                      
\usepackage{booktabs}                  
\usepackage{lipsum}                    
\usepackage{mwe}                       
\usepackage{tabularray}
\usepackage{float}
\usepackage{anyfontsize}
\usepackage{mdframed}

\begin{document}


\firstsection{Introduction}
\maketitle

The use of dashboards for information dissemination and decision-making is present in almost all companies, public or private organizations, profit or non-profit, spanning various industry sectors. As highlighted by Sarikaya et al. \cite{Sarikaya2019}, dashboards have become key elements to support data-driven decision-making.

Although dashboards are popular and useful for data visualization, their effectiveness and reliability depend on the data quality they present. The reliability of dashboards is closely related to the curation of these data, which involves providing clear information about origin of the data and the visual components of the dashboard \cite{Shankar2021}. The research field that studies the origin of an digital or material object is known as provenance. According to Shankar et al. \cite{Shankar2021}, provenance provides details on how a given resource are collected, transformed, normalized, and made available to the user. The Word Wide Web Consortium (W3C) published in 2013 the W3C-PROV, a standard for provenance that can be used to describe how data was gathered to ensure it is utilized in a meaningful way, ascertain who owns an object and their rights over it, evaluate information to decide if it is reliable, check if the steps taken to achieve a result are in line with given specifications, and replicate how something was created. The W3C defines provenance as a record that describes people, organizations, entities, and activities that had a role in the production, influence, or delivery of a piece of data or an object \cite{Moreau2013}.

The emergence of the COVID-19 pandemic led universities, businesses, and governments in various countries around the world to create dashboards to disseminate pandemic indicators such as the number of individuals infected, deaths, hospitalizations, and later on vaccination metrics. According to Shankar et al. \cite{Shankar2021}, on the one hand, dashboards such as those provided by the Centers for Disease Control and Prevention \cite{COVID_DATA_Tracker_2022} and John Hopkins University \cite{JHU2022} provided detailed information on the provenance of the data. On the other hand, other dashboards were questioned due to a breakdown in the data provenance chain. The author mentions the notable case that occurred with the state of Florida COVID-19 dashboard \cite{COVID-Florida}, where data scientist Rebekah Jones was allegedly fired for refusing to manipulate infection rate data to expedite economic reopening plans.

The Infovis for Public Health Research Project, conducted between 2020 and 2022, with funding from the Pan American Health Organization (OPAS), in response to the demand of the Department of Monitoring and Evaluation of the Brazilian Unified Health System - Executive Secretariat of the Ministry of Health (DEMAS/SE-MS - \textit{Departamento de Monitoramento e Avaliação do SUS - Secretaria Executiva do Ministério da Saúde do Brasil}), evaluated the experience of public health managers with the dashboards provided by the Strategic Management Support Room (SAGE - \textit{Sala de Apoio à Gestão Estratégica}). Relevant issues regarding the reliability of the information displayed on these dashboards were identified. During the interviews conducted by the project team, concerns were raised about the lack of information regarding the data source, the absence of identification of the responsible party for the information presented, as well as the need for clarity regarding the frequency and periodicity of data updating - common pieces of information in data provenance.

The term "provenance" originated in the fine arts \cite{Shankar2021, ProvenanceResearchToday, Fuhrmeister2019, PROV-FINE-ART-2019, Gramlich2017, Cheney2009, Moreau2006}, where it refers to information about the creation, ownership history, custody chain, modifications, or influences suffered by an artwork or manuscript.

In computing, the notion of provenance has been adopted and extended to work with digital artifacts, mainly applied to concepts such as data and user interaction, cognition, visualization, and visual analysis. In computing, provenance goes beyond the origin or history and includes the transformation process and other contextual information \cite{Shankar2021}.

Provenance is inherently context-dependent as different domains exhibit unique requirements and utilize provenance in diverse ways. It is essential to understand the specific problems that provenance addresses within each context and tailor its implementation accordingly \cite{Cheney2009}.

In an effort to create a provenance model that is comprehensive enough to accommodate different provenance use cases, the World Wide Web Consortium (W3C) created the W3C-PROV provenance model in 2013. This model has become a de facto standard, with research and applications conducted in various domains \cite{ProvenanceDimensions}. The W3C-PROV model defines provenance as information about entities, activities, and people who produce a given thing or data. It can be used to evaluate its quality and reliability \cite{Zhang2020, PROV-Overview}.

Recent studies have highlighted the need for further investigation into provenance in dashboards. Zhang et al. \cite{Zhang2022} highlight the credibility crisis that emerged amid the COVID-19 pandemic. On the one hand, the public seeks more reliable information, while on the other hand, government agencies, health experts, and dashboard designers express reservations about the reliability of their multiple sources, which could lead to confusion and even undermine trust in the authorities. Shankar et al. \cite{Shankar2021} highlighted the issue of trustworthiness surrounding certain COVID-19 dashboards, which comes under scrutiny when the provenance chain between the data and the dashboard was disrupted. Sarikaia et al. \cite{Sarikaya2019} emphasize the deficiency in provenance information, particularly within social organizational domains and urban informatics. In these contexts, data quality, trust, and accountability are of the utmost importance, yet current dashboards frequently fall short in eliciting these aspects, despite their pivotal role in enhancing public trust. Matheus et al. \cite{Matheus2020} conducted a study aimed at understanding and facilitating the design of dashboards to promote transparency and accountability. The research carried out by Ivankovic et al. \cite{Ivankovic2021}, which analyzed various dashboards related to COVID-19, revealed that 34 of the 158 dashboards studied (24.7\%) did not provide information on the data source. Jarske et al. \cite{Jarske2022} proposed a framework designed to facilitate effective communication of practical data provenance, while also providing direction for forthcoming advancements in provenance visualization techniques tailored specifically for dashboards.

Concerns about the reliability of the data available on the Web are almost as old as the Web itself. The invention of the Web occurred in 1989, the first graphical Web browser, Mosaic, was created in 1993, and significant Web expansion began in 1995 \cite{Brugger}. As early as 1997, the Web inventor, Tim Berners-Lee, foreseeing the issues of reliability in Web applications, suggested that Web applications should provide a button, which he referred to as the "Oh, yeah?" button, for users to press whenever they lost trust in the information provided. When pressed, the system would display a set of metadata and annotations that would allow the user to regain confidence \cite{Moreau2010}.

Motivated by the absence of a standard for visualizing provenance metadata in dashboards and inspired by the pioneering vision of Tim Berners-Lee, this paper aims to propose a representation model that allows the standardization, modeling, generation, capture, and visualization of the dashboard provenance. 

The authors of this paper anticipate that a standardized model for representing dashboard provenance will be beneficial by assisting architects and developers in constructing more effective and user-friendly provenance visualization interfaces; and enabling the end-users of dashboards to access comprehensive and reliable provenance information about the data and visualizations presented on the dashboard. This improved understanding should empower users to trust the data and make well-informed decisions based on the provenance information.

By adopting a standardized model, the visualization of provenance should become more consistent and coherent, making it easier for users to comprehend and interpret the origins and transformations of the displayed data and visualizations. A standardized model encourages interoperability and uniformity between different dashboard applications, which enhances collaboration and makes it easier to share provenance data between different platforms.

\section{The need for a Dashboard Provenance Model}

Just as different knowledge domains have specific applications and need for provenance, dashboards also have their specificities. In this section, we present a dashboard provenance model.

The proposed model extends the W3C-PROV model to address the particular needs of dashboards, taking into account the common entities, collections of entities, activities, and agents involved in the typical lifecycle of a dashboard from its demand, construction, operation, and maintenance. This was identified during the Infovis for Public Health Research Project and by examining the COVID-19 dashboards chosen in the study conducted by Ivankovic et al. \cite{Ivankovic2021}.

The W3C-PROV offers a generic provenance data model that can be tailored and implemented in a wide range of domains, demonstrated by the 38 use cases proposed by the Provenance Incubator Group\cite{ProvenanceUseCases}. Other examples of applications include exchange in healthcare \cite{HL7-International}, educational resources \cite{ProvOER2023}, geoprocessing workflows \cite{Zhang2020} and internet art \cite{PROV-FINE-ART-2019}. Although each dashboard developer can use W3C-PROV to create their own solutions to display provenance information to their users, our research is focused on providing a standardized model that can be easily applied to any dashboard.

A standardized model for dashboard provenance should furnish the relationships and hierarchies to the entities within the dashboard (both visual and data), outline the customary activities implicated, and delineate the roles typically undertaken by agents throughout the lifecycle of a dashboard.

As stated by Ragan et al. \cite{Ragan2016}, there exist five different types of provenance information: i) data provenance: encompasses the lineage of alterations and movements of data, which may encompass activities like data subsetting, merging, formatting, transformations, and simulations aimed at generating or incorporating new data; ii) visualization provenance: pertains to the sequence of graphical views and states of visualizations; iii) interaction provenance: involves a record of user actions and commands executed within a system; iv) insight provenance: captures the progression of cognitive outcomes and information attained from the analysis process, including analytical findings, hypotheses, and deductions; and v) rationale provenance: encompasses the chronicle of reasoning and intentions behind decisions, hypotheses, and interactions. This work will focus on data and visualization provenance that includes information about data sources, preprocessing steps, and design choices made during the creation of the visual element. 
However, it is important to highlight that this study will not focus on the visualization provenance resulting from user interaction with the dashboards, including the record of visualization states.

The construction of a dashboard provenance model aims to provide a consistent and comprehensive model to represent the origin and history of the different components that make up dashboards, enabling the creation of provenance visualizations that facilitate a clear and accurate understanding of the context in which a particular dashboard was created.

In the early days of W3C-PROV development, the Provenance Incubator Group provided a list of provenance dimensions to guide the design of use case scenarios \cite{ProvenanceDimensions}. The provenance dimensions were grouped into content, management, and use of provenance. The content of provenance leads to the meaning of provenance in a given context, the modeling of provenance identifying its basic elements (entities, agents, and activities), relationships and properties among the basic elements, evolution and versioning, justifications for decisions, and expected results. The management of provenance leads to issues related to publication (visualization), access (browsing and querying), dissemination control (access control, licensing, and legal issues), and scale (volume) of provenance content. Finally, the use of provenance leads to issues related to the user experience with the provenance visualization.

The design of use-case scenarios for the dashboard provenance can vary depending on the goals of the dashboards and the user-specific needs. Using a provenance model for dashboards can improve the design of provenance, ensuring that it aligns with the provenance dimensions, particularly to the achievement of the user needs and the dashboard's goals.

The proposed model to represent the provenance of dashboards aims to provide a set of metadata that allows users to evaluate the reliability, quality, and consistency of the content displayed on dashboards that includes visual and data artifacts.


A dashboard can be interpreted as a collection of visual entities and data, agents, and activities. Visual entities can include charts, maps, tables, infographics, or text. Data entities can be provided in different formats such as CSV (Comma Separated Value) \cite{CSV2005}, JSON (JavaScript Object Notation) \cite{JSON2017}, GeoJSON (JavaScript Object Notation-based geospatial data interchange format) \cite{GEOJSON}, among other formats commonly used on the Web. It is necessary to map all individual entities and those that are part of a visual and data artifact collection that make up the dashboard for a comprehensive representation of provenance. Agents can be organizations, software systems, or individuals responsible for creating, maintaining, and updating the dashboard. Activities represent the actions or operations performed by agents within the dashboard's lifecycle, such as creating, modifying, or updating visual components, data entities, or the dashboard itself. 

The agents responsible for the dashboard can be categorized as follows:

\begin{itemize}
\item Demand-side agents: these are the agents (organization, department, or people) who initiated the development of the dashboard. They represent the stakeholders who identified the need for the dashboard to support their specific requirements and goals;

\item Development and maintenance agents: these agents are responsible for creating, designing, and maintaining the visual entity and data in the dashboard. They can be individuals or teams within the organization or external contractors who handle the technical aspects of the dashboard development and ensure its operation;

\item Curation agents: these agents are responsible for curating and managing the visual and data entities displayed in the dashboard. They ensure quality, accuracy, and reliability;

\item Data update agents: these agents are responsible for regularly updating the data presented on the dashboard. They can be data scientists, analysts, database administrators, or automated processes that retrieve new data, perform data transformations, and incorporate updated information into the dashboard.
\end{itemize}

The activities involved in creating, maintaining, and updating the dashboard include the following:

\begin{itemize}
	\item Creation and maintenance: involves the design, development, and maintenance of the dashboard, including the selection of visual elements, data sources, and overall layout. It encompasses tasks such as collecting and organizing data, creating visualizations, and integrating various components into a cohesive dashboard;

	\item Sustainability: ongoing efforts and practices to maintain and preserve the stability, viability, and longevity of the dashboard, visual entities, and data. Sustainability activities include regular monitoring and evaluation of the dashboard's performance, gathering user feedback, and incorporating it into future updates; addressing technical issues and ensuring compatibility with new technologies or data sources, and aligning the dashboard with evolving user needs and organizational goals;

	\item Data update: data update activity refers to updating the data used in a dashboard. It involves retrieving new data, incorporating them into the existing dataset, and ensuring that the dashboard reflects current information.
\end{itemize}

By documenting the above activities, it becomes possible to comprehend the development and evolution of the dashboard over time.

To identify basic provenance related to visual elements, it is proposed to divide the dashboard into three visual layers: i) the dashboard as a whole; ii) the dashboard subtopics; and iii) individual visual entities. Data entities can be associated with one or more of these visual layers.

The division into layers aims to provide users with an understanding of provenance at different levels of detail, allowing for the display of provenance information for individual visual entities and their relationships with other visual entities grouped under a common theme and with the dashboard as a whole. This approach enables the provision of different levels of provenance to users, as they can access more detailed or consolidated information based on their interests.

\begin{figure}[htpb]
	\centering
	\label{fig:ProvenanceLayer}
	\begin{subfigure}[b]{0.977\columnwidth}
		\centering
		\includegraphics[width=\textwidth]{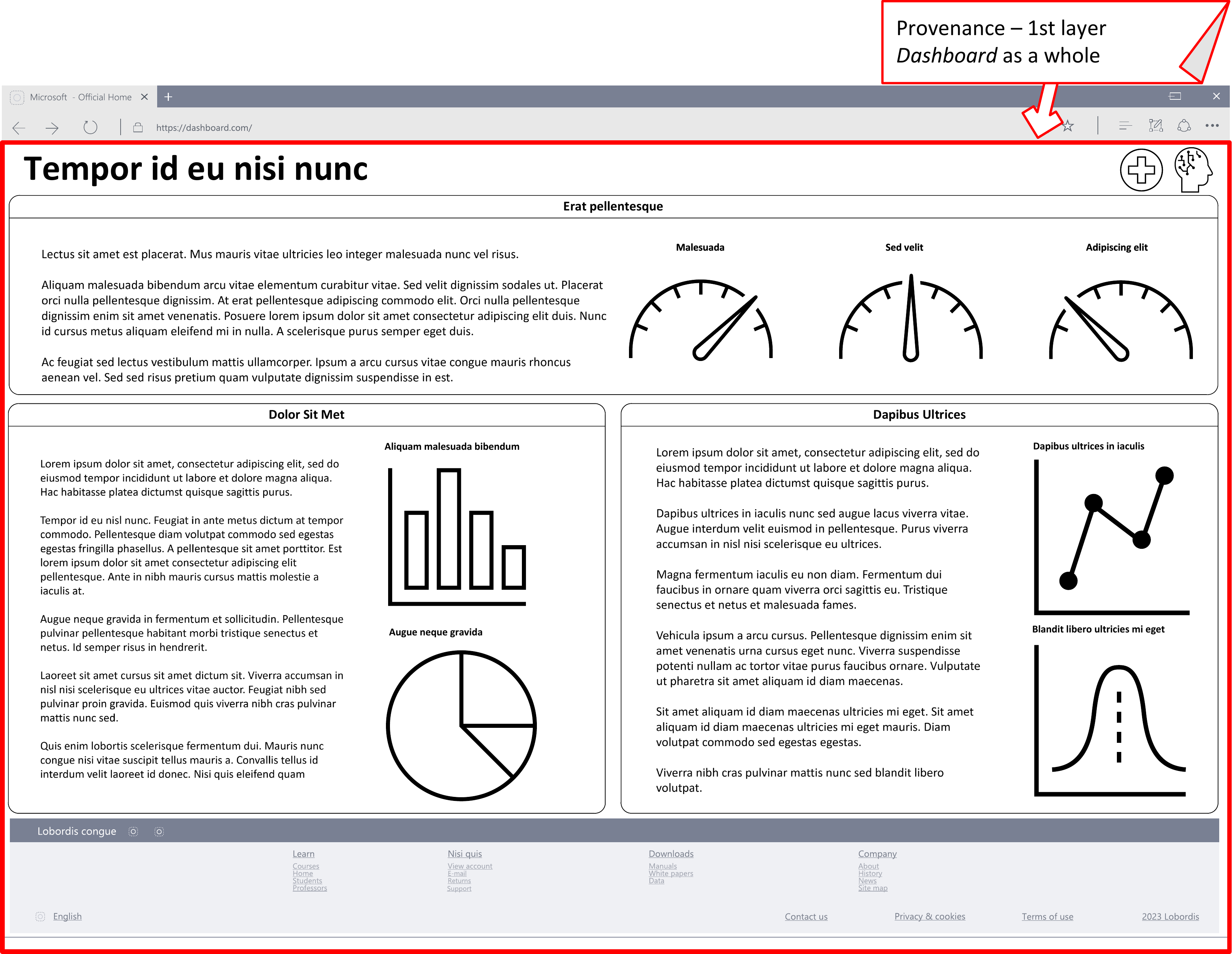}
		\caption{Dashboard provenance - first layer.}
		\label{fig:ProvenanceLayer1}
	\end{subfigure}%
	\\%
	\begin{subfigure}[b]{0.977\columnwidth}
		\centering
		\includegraphics[width=\textwidth]{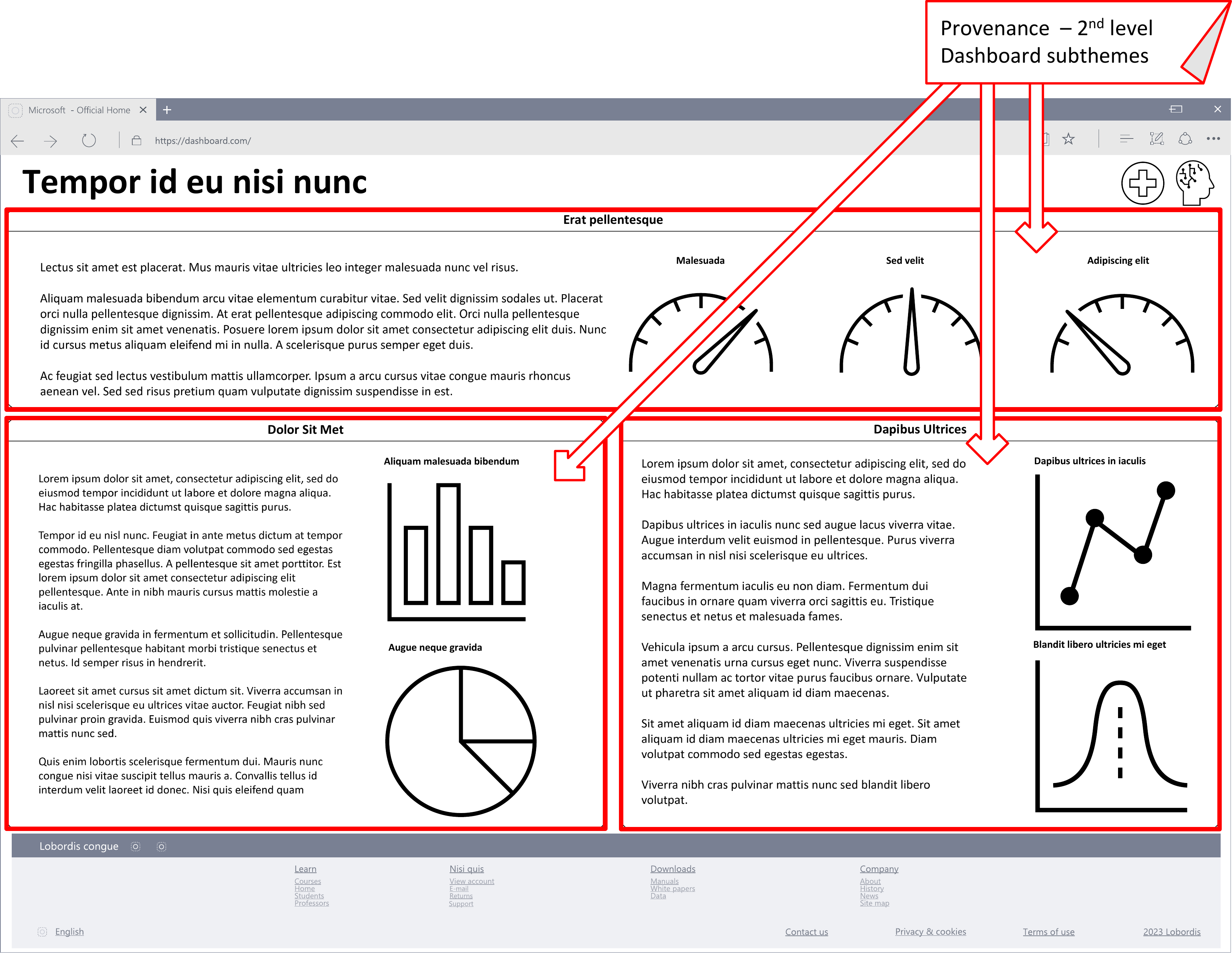}
		\caption{Dashboard provenance - second layer.}
		\label{fig:ProvenanceLayer2}
	\end{subfigure}%
	\\%
	\begin{subfigure}[b]{0.977\columnwidth}
		\centering
		\includegraphics[width=\textwidth]{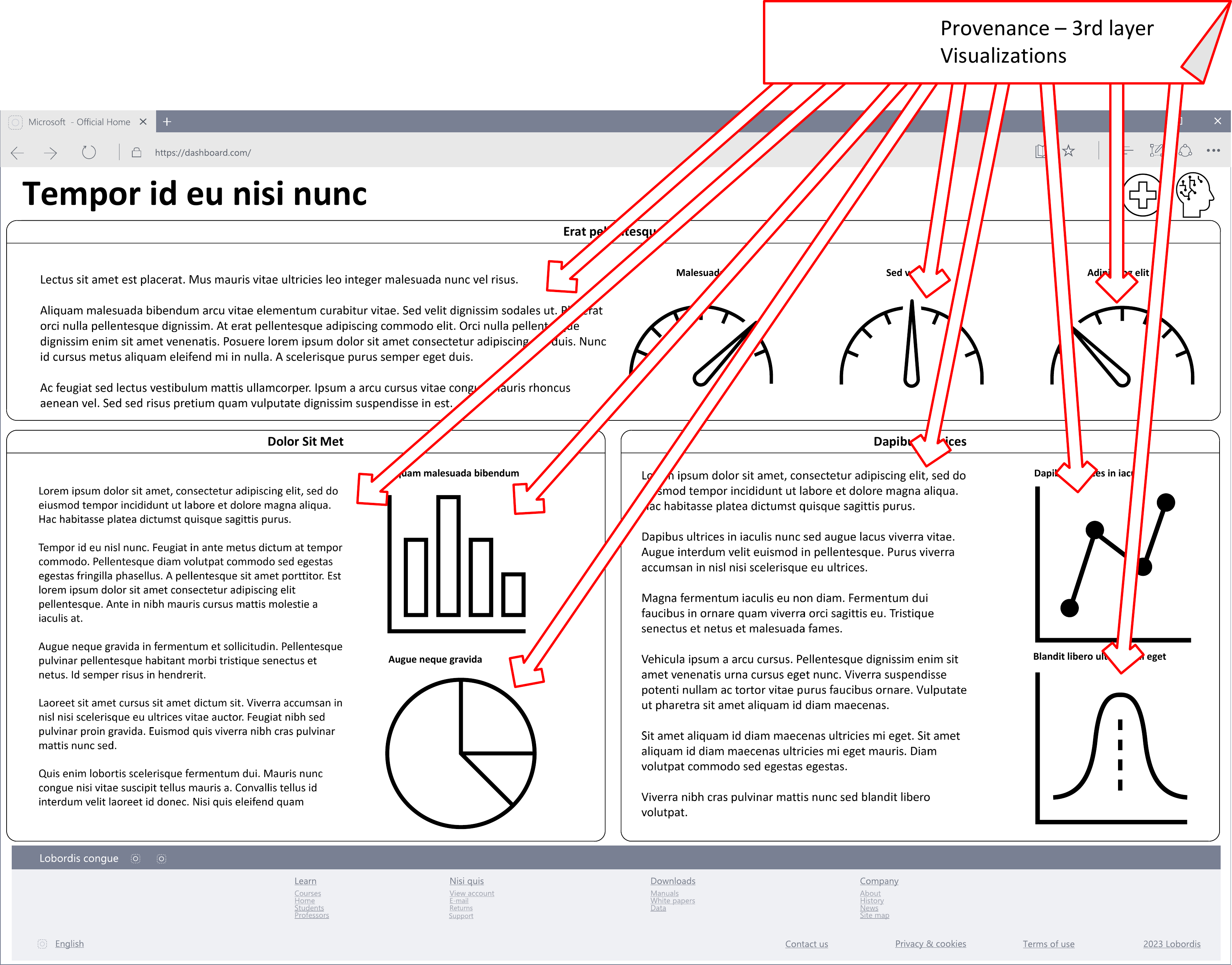}
		\caption{Dashboard provenance - third layer.\\}
		\label{fig:ProvenanceLayer3}
	\end{subfigure}%
	\label{fig:ProvenanceLayer}
	\caption{Illustrates how the proposed dashboard provenance can be structured in layers that can provide the user with varying levels of provenance information.}	
\end{figure}

\cref{fig:ProvenanceLayer1} represents the first layer of provenance, which encompasses the dashboard as a whole. At this level, the dashboard is considered a single entity and a collection of other entities. This layer serves to provide an overview of the dashboard and achieve the following objectives: i) describe the general properties of the entity, such as title, descriptive text, global references, annotations, version, and creation date; ii) present the agents responsible for the dashboard curation, creation, and maintenance, including name, uniform resource locator (URL), description, and annotations; iii) capture the activities involved in the creation and maintenance of the dashboard.

\cref{fig:ProvenanceLayer2} represents the second layer of provenance, which refers to the subtopics addressed within the dashboard. For example, in a dashboard related to COVID-19, distinct topics such as statistics on the number of cases, the number of deaths, vaccination, and associated comorbidities can be addressed. Similarly to the first layer of provenance, in the second layer, entities can be understood as unique entities or as a collection of entities related to a specific subtopic addressed within the dashboard. Each subtopic may involve different entities, agents, and activities for which the maintenance and display of provenance metadata is important. This allows the recording of properties and provenance information related to the addressed subtopics, the agents responsible for the curation, creation, maintenance, and updating of the second layer, and the activities performed.

For some dashboards, the definition of a second layer of provenance may not be applicable, as there may be no distinguishable subtopics that justify this division.

Finally, \cref{fig:ProvenanceLayer3} represents the third layer of provenance, which refers to individual entities such as graphs, tables, or explanatory text. Similar to the collections of entities, the maintenance of provenance metadata, agents, and performed activities is also necessary for individual entities.

\newpage

Regardless of the layer in which the basic elements of dashboard provenance are located, the following properties can be included:

\begin{itemize}
	\item Entities:
	\begin{itemize}
		\item Visual: name, identification, description, creation date, versioning, authorship, maintainer, ownership, source, data sources, metadata source, relationships, annotations, and explanations (that should include validity and quality measures, generation parameters, graph construction steps, how the data attributes are mapped to the graph elements, graph interactions, data associated with individual nodes and edges in the graph).
		\item Data: name,  identification, type, creation/maintenance date, frequency of update, version, authorship, ownership, maintainer, responsibility, source, metadata source, relationships, annotations, and explanations (which should include data collection methods, data processing steps, data transformations).
	\end{itemize}
	
	\item Activities: name,  identification, type, description, type, inputs, outputs, parameters, dependencies, start time, end time, frequency of execution, authorship, relationships, versioning, validation, and quality measures.
	
	\item Agents: name,  identification, role, contact information, description, responsibilities, affiliation, contributions, relationships, trustworthiness.
\end{itemize}

\begin{table}[H]
	\tiny
	\caption{Set of metadata for dashboard provenance.}
	\label{tab:MetadadosProveniencia}		
	
	\begin{subtable}[htpb]{1\columnwidth}
		\centering
		\caption{W3C-PROV - basic elements.}
		\begin{tabular}{p{14mm}p{30.5mm}p{30.5mm}}
		\hline
		\textbf{Metadata} & \textbf{Description} & \textbf{Application}\\
		\hline	
        prov:agent & Agents are the organizations, departments, and sectors responsible for the dashboard. &   To identify the agents involved in the development of the dashboard, as well as those responsible for the maintenance of visual entities and data, data and visual artifacts curation, and data updating activities. \\

		prov:entity & Entities can include the dashboard itself, its visual components, groups of visual components, and the data.  & To identify the visual and data entities. In dashboards, visual entities are derived from data entities or other visual entities. \\	
		
		prov:actity & On the dashboards, there are activities for the creation and maintenance of visual and data entities, and for data update. & The activities of development and maintenance of visual and data entities are important to understand when the entities were created and modified due to evolutionary or corrective maintenance. There may be a single activity for data updates or specific activities for each dataset in the dashboard. \\	
		\end{tabular}
		\label{tab:BasicElements}
	\end{subtable}
\end{table}

\begin{table}[H]
	\ContinuedFloat
	\tiny
	\begin{subtable}[htpb]{1\columnwidth}
		\centering
		\caption{W3C-PROV - basic relations.}
		\begin{tabular}{p{14mm}p{30.5mm}p{30.5mm}}
		\hline
		\textbf{Metadata} & \textbf{Description} & \textbf{Application}\\
		\hline	

        prov: \\wasGeneratedBy & Specifies which activity generated the entity. & Connects a specific entity to the activity that generated or transformed it. This can be an activity that developed or performed maintenance on a visual entity or updated a dataset. \\ 
		
		prov: \\wasDerivedFrom & Connects a visual entity to other visual or data entities that are its source or origin. & Inform the user which entities a given entity was derived from. \\
		
		prov: \\wasAttributedTo & Specifies which agent a given entity is assigned. & Inform the user to which agent a given entity has been assigned. \\
		
		prov:startedAtTime & Specifies when activity was started. & Notifies the user if a given activity was started. \\
		
		prov:endedAtTime & Specifies when a given activity was ended. & Informs the user if a given activity was ended.  \\
		
		prov:used & Specifies whether an existing entity was modified or a new one was generated by a given activity and the entity used. For example, in a dashboard, it can indicate that a specific dataset was used to update the dashboard data or that a new version of an entity was produced using an existing one. & Informs the user which entity was used to modify an existing entity or generate a new entity. \\
		
		prov: \\wasAssociatedWith & Specifies the agent responsible for executing an activity. & Informs the user about which agent (software or organization) is responsible for updating a dataset or the agent responsible for maintaining a specific visual component. \\
		
		prov: \\actedOnBehalfOf & Specifies hierarchical relationships or delegated responsibility between organizations and departments. & Displays hierarchical or delegation of responsibility among agents, giving users an understanding of the relationship between them. \\
		\end{tabular}
		\label{tab:BasicRelations}
	\end{subtable}
\end{table}

\begin{table}[H]
	\ContinuedFloat
	\tiny
	\begin{subtable}[htpb]{1\columnwidth}
		\centering
		\caption{W3C-PROV - extended relations.}
		\begin{tabular}{p{14mm}p{30.5mm}p{30.5mm}}
			\hline
			\textbf{Metadata} & \textbf{Description} & \textbf{Application}\\
			\hline	

	    prov:generatedAtTime & Specifies the time at which the entity was generated. & Informs the user about the generation time of an entity. It can indicate, for example, the date and time when a particular dataset was last updated. This information clarifies the freshness and accuracy of the data within the entity, allowing users to make informed decisions based on the latest available information. \\
		
		prov:collection & Specifies an entity composed of a set of visual and data entities. & The statement that an entity consists of a set of entities is valuable for establishing relationships between them, because it means that they share common data sets or related information. \\
		
		prov:softwareAgent & Specifies that a given agent is a software component. & This property explicitly states that the entity responsible for executing an activity is a software component. For example, it can indicate that a software program performs daily updates on a dataset. \\
		
		prov:organization & Specifies that a given agent is an organization. & It can be used to indicate that a specific agent responsible for the dashboard or dataset is an organization. \\
		
		prov:hadPrimarySource & Specifies that a given entity was derived from a primary source. & Informs the user that a specific entity has a primary data source. Help users understand the origin and reliability of the data presented in the entity. \\
		
		prov:type & Specifies if a given agent is an organization, department, person, or software. & The user is informed about the type of agent involved in the activity or responsible for the entity. \\
		\end{tabular}
		\label{tab:ExtendedRelations}
	\end{subtable}

\end{table}	
	
\begin{table}[H]
	\ContinuedFloat
	\tiny
	\begin{subtable}[h]{1\columnwidth}
		\centering
		\caption{Dashboard provenance model - extended properties.}
		\begin{tabular}{p{14mm}p{30.5mm}p{30.5mm}}
			\hline
			\textbf{Metadata} & \textbf{Description} & \textbf{Application}\\
			\hline	
		dash:name & Name assigned to the provenance element (entity, agent or activiy), which will be visible to the user.  &  The user can identity the provenance element by their name that could be the same of the title of a graph (visual entity) or the name of the dashboard publisher (agent). \\
		
		dash:version & Identification mechanism, particularly significant when there are multiple iterations or updates throughout the dashboard lifecycle. & A version number provides traceability, transparency, compatibility and facilitating effective communication regarding different dashboard versions.  \\
		
		dash:role & Describing the role or function of the agent in the overall process, such as demand-side agent, development and maintenance agent, data curation agent, data update agent. & By knowing the role of an agent, the user can better understand the role and responsibilities played by an agent in the dashboard context, which conveys credibility and transparency. \\
		
		dash:description & A textual description that provides additional information about the provenance element (entity, agent or activity), including its purpose, content, or context. &  The additional information provided to the user in a dashboard may contain important information to contextualize the user in relation to the topic addressed, providing clarity and understanding.\\
		
		dash:annotations & Relevant comments or observations associated with the provenance element, providing additional details or clarification on its nature or content. This includes the definition of technical terms or specific concepts, the explanation of measurement units, the description of formulas or calculations used, among others. Annotations can be interpreted of a extended version of descriptions. & This information helps the user to correctly interpret the visualizations or data and avoid misunderstandings or misinterpretations. Annotations could also supply information about agents (rules, responsibilities, contributions and contact information) and activities (inputs, outputs, parameters, dependencies, validation, and quality measures)\\
		
		dash:contactInformation & Providing contact details, such as email address, phone number or organizational affiliation & Contact information allow users to reach out to the agent for further information or inquiries.   \\
		
		dash:trustworthiness & Any certification, credential, or trust metrics associated with the agent. & This information assures users of the agent's reliability and expertise in the field. \\
		
		dash:url & The url refers to the Web address where users can access additional details or resources related to the entities, agents and activities present in the dashboard. & By including the url information about entities, agents and activities, users can access and verify the sources and background information of the provenance resources displayed in the dashboard, enhancing their understanding and trust in the data and visualizations presented. \\
		
	\end{tabular}
	\label{tab:DashExtendedRelations}
\end{subtable}	
	\caption*{Source: provided by the authors based on the specifications of W3C-PROV.}
\end{table}

The \cref{tab:MetadadosProveniencia} showcases a compilation of provenance metadata as defined by W3C-PROV \cite{PROV-Overview}, along with the additional properties tailored for dashboard provenance. The \cref{tab:BasicElements} provides a detailed exposition of the fundamental constituents within W3C-PROV. Furthermore, \cref{tab:BasicRelations} and  \cref{tab:ExtendedRelations} respectively outline the elementary and expanded relations of W3C-PROV that are foreseen to be employed within the realm of dashboard provenance. The \cref{tab:DashExtendedRelations} presents the collection of custom metadata, under the prefix \textbf{dash}, which expands the W3C-PROV framework to meet the requisites of dashboard provenance. This extension enriches the W3C-PROV  with additional context and provenance details.

\section{Applying the dashboard provenance model} 

This modeling process sample uses a preexisting dashboard, although it is highly recommended to design the provenance of the dashboard parallel to the dashboard design.

\begin{figure}[H]
	\centering
	\begin{subfigure}[b]{0.86\columnwidth}
		\centering
		\includegraphics[width=\textwidth]{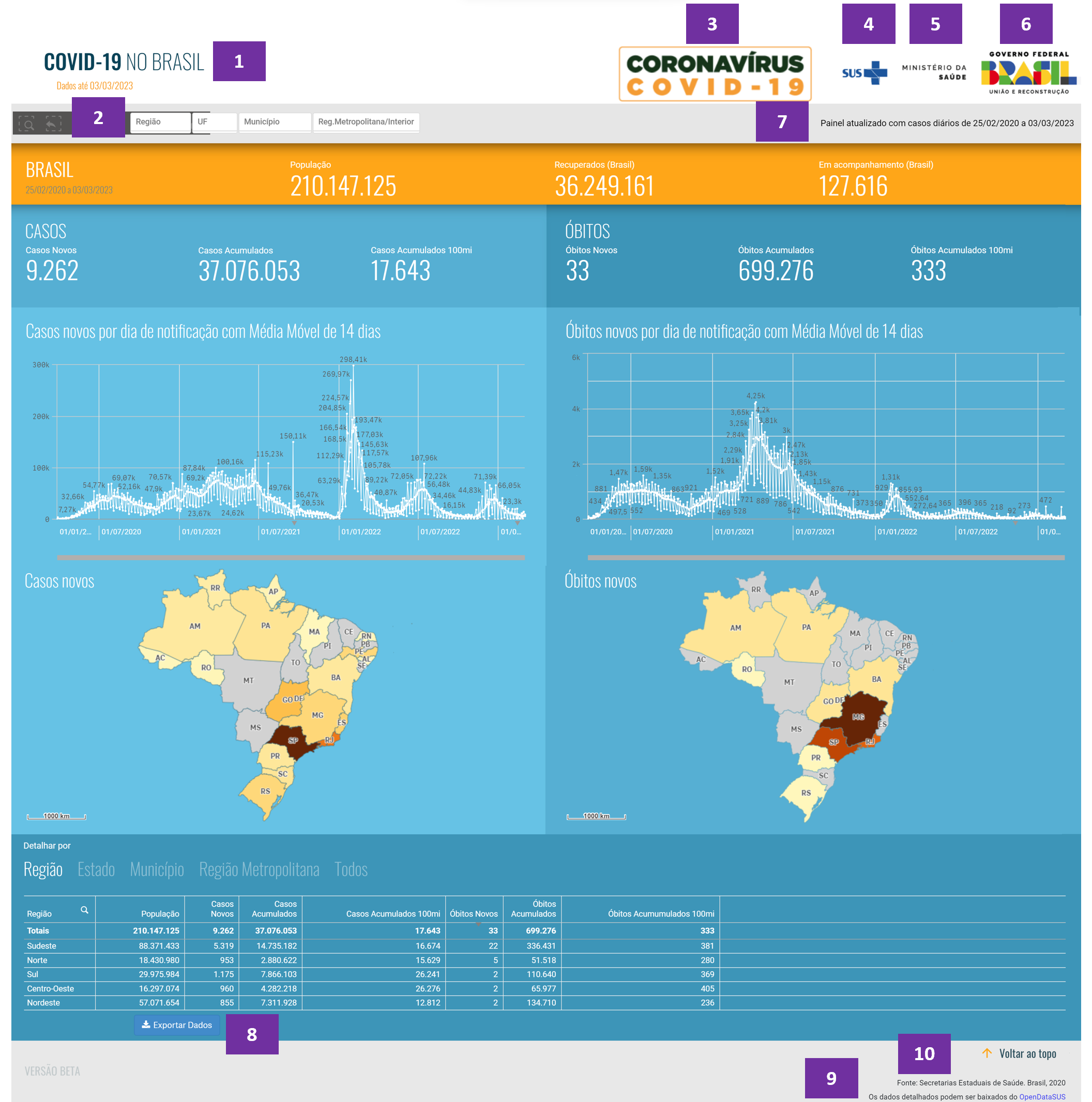}
		\caption{Dashboard COVID-19 no Brasil - first layer.}
		\label{fig:COVID-19-Brasil-1}
	\end{subfigure}%
	\label{fig:COVID-19-Brasil}
	\\%
%
	\begin{subfigure}[b]{0.86\columnwidth}
		\centering
		\includegraphics[width=\textwidth]{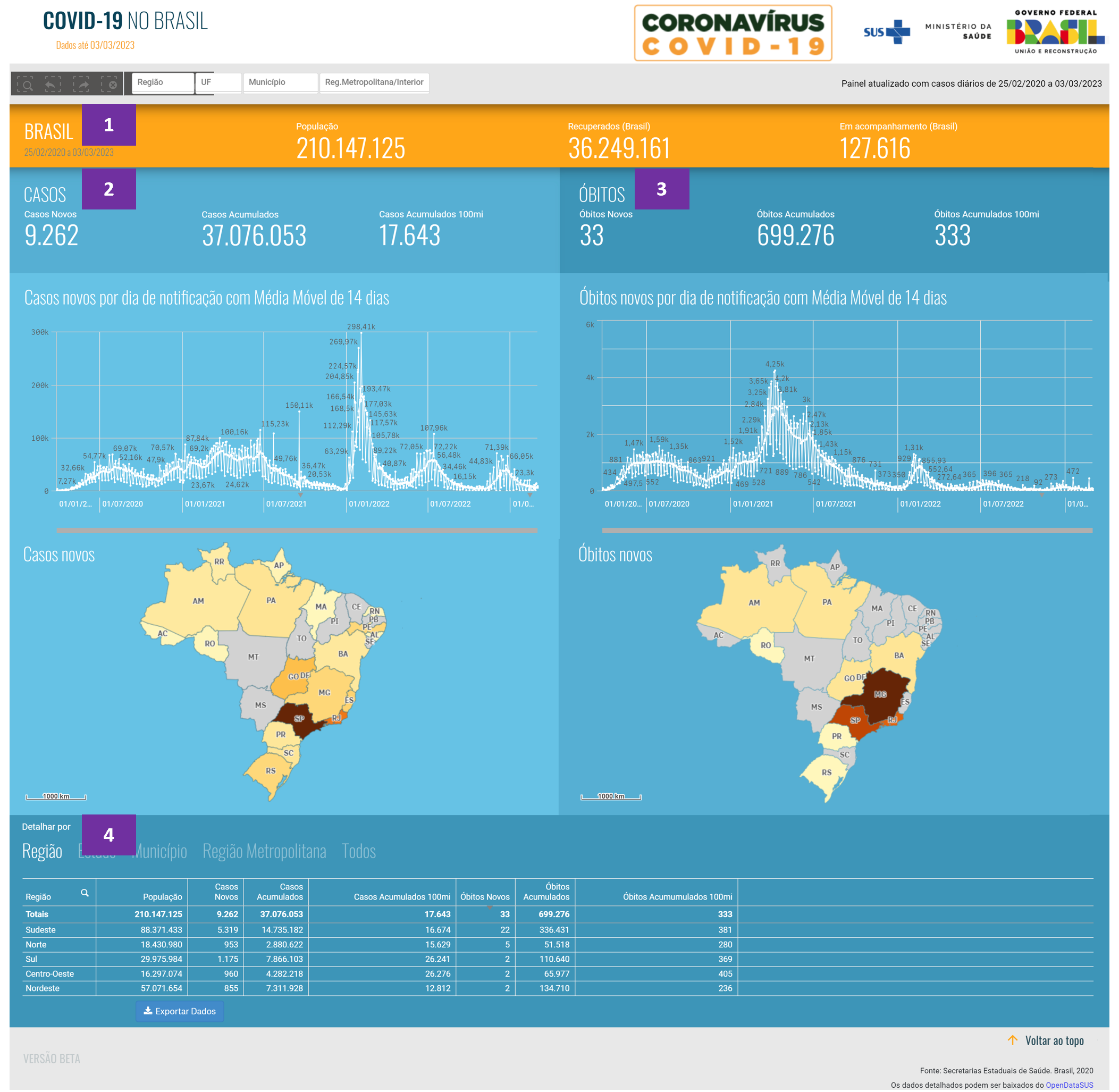}
		\caption{Dashboard COVID-19 no Brasil - second layer..}
		\label{fig:COVID-19-Brasil-2}
	\end{subfigure}%
	\\%
%
	\begin{subfigure}[b]{0.86\columnwidth}
		\centering
		\includegraphics[width=\textwidth]{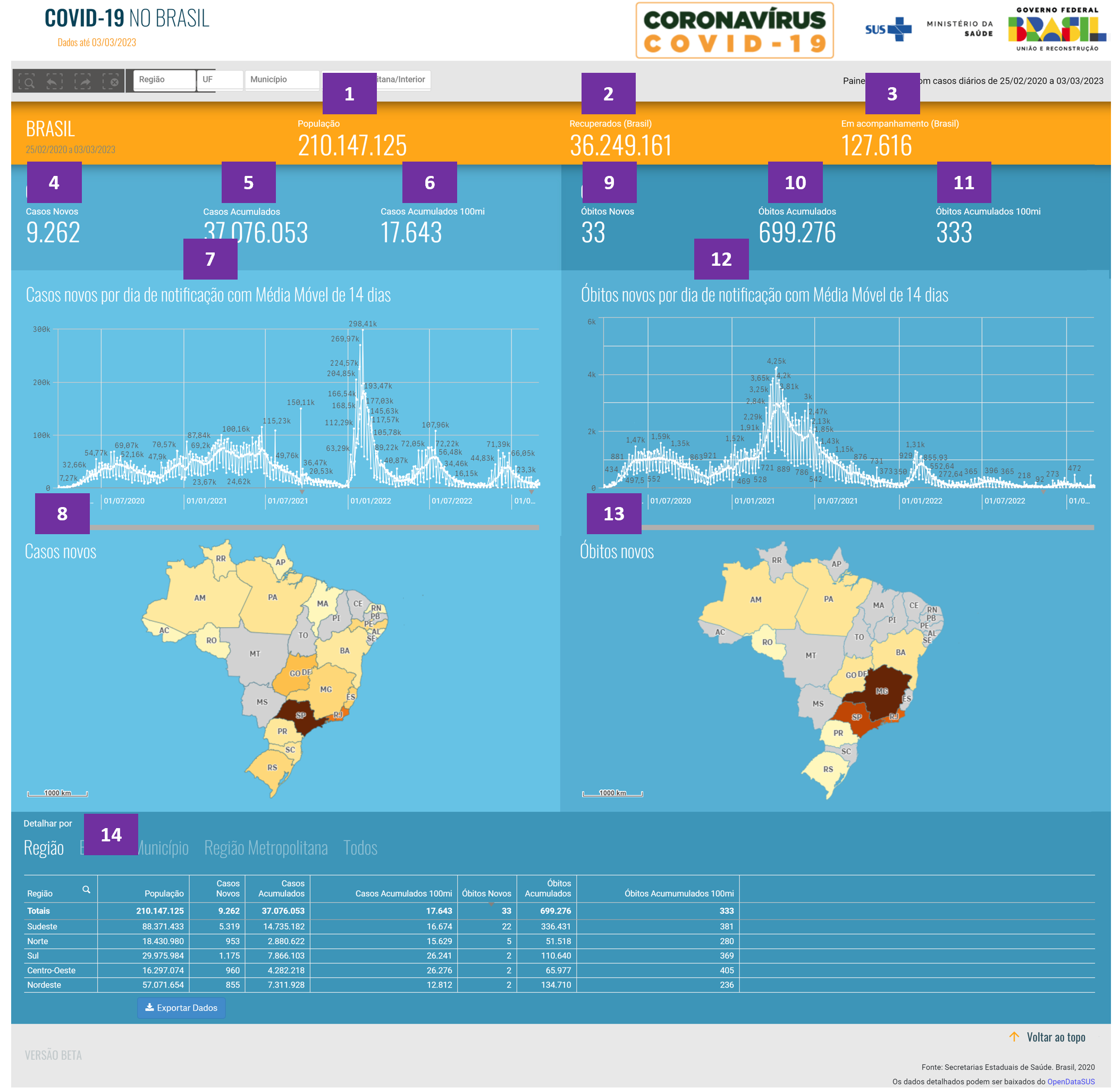}
		\caption{Dashboard COVID-19 no Brasil - third layer.}
		\label{fig:COVID-19-Brasil-3}
	\end{subfigure}%
	\subfigsCaption{Displays the hierarchical structure of the three proposed layers of dashboard provenance employed in the \textit{COVID-19 no Brasil} context \cite{COVID-19-Brasil}.}
	\label{fig:COVID-19-Brasil}
\end{figure}

On the first level of provenance, the dashboard as a whole, shown in \cref{fig:COVID-19-Brasil-1}, a set of related provenance metatada was identified: 1) the name of the dashboard; 2) the last date covered by the dashboard (data until 2023-03-03); 3) the logo used by the Brazilian government to represent the COVID-19 taskforce; 4) the logo of the Unified Health System (SUS - \textit{Sistema Único de Saúde}); 5) the logo of the Brazilian Ministry of Health (\textit{Ministério da Saúde}); 6) the logo of the Brazilian Federal Government (\textit{Governo Federal do Brasil}); 7) the interval of data displayed on the dashboard - Dashboard updated with daily cases from 2020-02-25 to 2023-03-03; 8) download the data; 9) the data source - Brazilian State Departments of Health (\textit{Secretarias Estaduais de Saúde. Brasil}) - collection of agents; and 10) the curator of the data,
DataSUS (IT department of the Brazilian Unified Health System), the agent responsible for the curation of the data. Items 3, 4, 5, and 6 represent a hierarchy of agents: 3 acted on behalf of 4, that acted on behalf of 5, that acted on behalf of 6.

On the second level of provenance, shown in \cref{fig:COVID-19-Brasil-2}, the following visual subthemes were identified: 1) \textit{Brasil} - general data about COVID-19 in Brazil; 2) \textit{Casos} - number of infected people; 3) Óbitos - number of deaths; and 4) \textit{Detalhar por} - detailed number of cases. 

On the third level of provenance, shown in \cref{fig:COVID-19-Brasil-3}, the following visual entities were identified: 1) \textit{População} - Brazilian population card; 2) \textit{Recuperados (Brasil)} - Brazilian number of recovered cases card; 3) \textit{Em acompanhamento (Brasil)} - Brazilian number of monitored cases; 4) \textit{Casos novos} - new cases; 5) \textit{Casos acumulados} - number of accumulated cases; 6) \textit{Casos acumulados 100mi} - number of accumulated cases per 100,000 inhabitants; 7) \textit{Casos novos por dia de notificação com media móvel de 14 dias} - new cases per day of notification with 14-day moving average graph; 8) \textit{Casos novos} - new cases on a Brazilian choroplet map; 9) \textit{Óbitos novos} - new cases of deaths; and 10) \textit{Óbitos acumulados} - number of accumulated cases of deaths; 

\begin{figure}[htpb]
	\centering
	\includegraphics[width= 1\columnwidth]{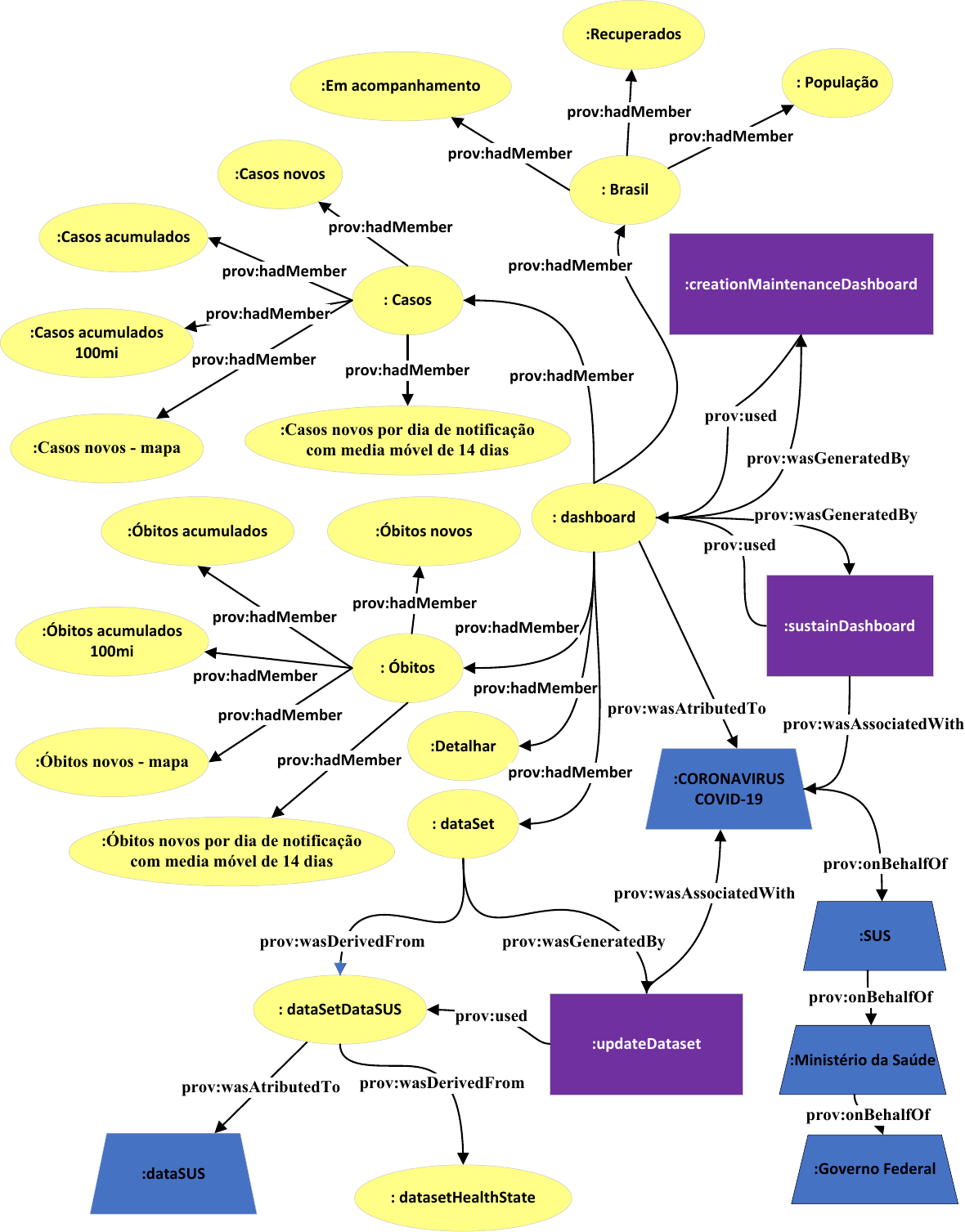}
	\caption{Provenance resources of the COVID-19 Brasil dashboard.}	
	\label{fig:COVID-BRASIL-PROV}
\end{figure}
\noindent 11) \textit{Casos acumulados 100mi} - number of accumulated cases of death per 100,000 inhabitants;  12) \textit{Óbitos novos por dia de notificação com media móvel de 14 dias} - new cases of death per day of notification with 14-day moving average graph; 13) \textit{Óbitos novos} - new cases of death on a Brazilian choroplet map; 14) \textit{Detalhar por} - Table showing COVID-19 indicators in Brazil at different aggregation levels.

The \textit{COVID-19 no Brasil} dashboard \cite{COVID-19-Brasil} is depicted in \cref{fig:COVID-BRASIL-PROV}, which shows the basic provenance structure of the dashboard, including its basic elements (activities, entities and agents) and its basic and extended relationships.

The following sample of provenance ontology exemplifies how the proposed dashboard provenance model extends the W3C-PROV and provides metadata to describe the provenance of the COVID-19 Brasil dashboard \cite{COVID-19-Brasil}.

\newmdenv[linecolor=gray!10,frametitle=Example]{Example} \begin{Example}[backgroundcolor=gray!10]
@prefix xsd:  <http://www.w3.org/2001/XMLSchema\#> .\\
@prefix dash: <http://dash.com/dash\#> .\\
@prefix prov: <http://www.w3.org/ns/prov\#> .\\
\\
:dashboard\\
a prov:Entity, prov:Collection;\\
dash:Name "COVID-19 NO BRASIL";\\
dash:Version Beta\\
dash:Description "Painel atualizado com casos diários de 25/02/2020 a 03/03/2023"\\
dash:annotations "Em atendimento a definição clínica oficial da condição pós-Covid-19 pela Organização Mundial de Saúde (OMS)."\\
dash:url "https://infoms.saude.gov.br/extensions/covid-19.html"\\
prov:wasGeneratedBy  :creationMaintenanceDashboard\\
prov:wasDerivedFrom  :dataset;\\
prov:wasAttributedTo :CORONAVIRUS COVID-19;\\
prov:hadMember: "Brasil", "Casos", "Óbitos", "Brasil", "Detalhar";\\
.
\\
:Casos\\
a prov:Entity, prov:Collection;\\
dash:Name "Casos";\\
dash:"1.0.0.0"\\
dash:Description "Casos novos por dia, por dia de notificação."\\
dash:annotations "Os casos novos são calculados a partir da diferença entre os casos informados da semana atual menos os casos acumulados da semana epidemiológica anterior."\\
dash:url "https://infoms.saude.gov.br/extensions/covid-19.html"\\
prov:wasGeneratedBy  :creationMaintenanceDashboard\\
prov:wasDerivedFrom  :dataset;\\
prov:wasAttributedTo :CORONAVIRUS COVID-19;\\
prov:hadMember: "Casos novos por dia, por dia de notificação com média móvel de 14 dias", "Casos novos",\\
.
\\
:CORONAVIRUS COVID-19 \#COVID-19 taskforce\\
a prov:Agent;\\
dash:Name "CORONAVIRUS COVID-19 taskforce";\\
dash:Role "Developer", "Maintenance;"\\
dash:Description "Força-tarefa criada para acompanhar a evolução da pandemia do COVID-19 no Brasil."\\
dash:trustworthiness "gov.br"\\
dash:url "https://saude.gov.br/covid-19/covid-19-taskforce.html"\\
prov:actedOnBehalfOf :SUS;\\
.
\\
:SUS \\
a prov:Agent;\\
dash:Name "Sistema Único de Saúde";\
dash:Role "demanding";\\
dash:Description "Sistema Único de Saúde é a denominação do sistema público de saúde brasileiro criado pela Constituição Federal de 1988."\\
dash:annotations "Em todo o país, o SUS deve ter a mesma doutrina e a mesma forma de organização, sendo que é definido como único na Constituição um conjunto de elementos doutrinários e de organização do sistema de saúde, os princípios da universalização, da eqüidade, da integralidade, da descentralização e da participação popular."\\
dash:contactInformation "Esplanada dos Ministérios - Bloco G, Edifício Sede - CEP: 70058-900 - Brasília/DF"\\
dash:trustworthiness "gov.br"\\
dash:url "https://www.gov.br/saude/pt-br"\\
.
\end{Example}

\section{Discussion and Future Work}

This paper addresses the need for a dashboard provenance model to provide comprehensive information about the visual and data entities within a dashboard. Additionally, we have emphasized the importance of identifying the agents involved in demanding, developing, and maintaining the dashboard, as well as those responsible for tasks such as data curation and ongoing support to ensure the sustainability of the dashboard.  

A key contributions of this work is the development of a dashboard provenance model. This model provides a structured approach to capturing and representing the provenance of dashboards, including the entities, agents, activities, and their properties and relationships. The next steps of this ongoing project must include:

\begin{itemize}
\item \textbf{Development of a formal data model for dashboard provenance}: After defining the main components of the provenance of the dashboard, it is important to define a data model that represents the structure and organization of the provenance data associated with the dashboard. Based on the W3C-PROV-DM \cite{PROV-DM}, the data model should specify how different elements of provenance, such as agents, activities, entities, their relationships, and properties, are represented and interconnected within the the dashboard. The data model should provide a comprehensive and standardized framework for storing, querying, and analyzing provenance data. As this study focuses on Web-based dashboards, a PROV-JSON \cite{werner2020provviz, Zhai2017, PROV-JSON} format should be considered.

\item \textbf{Verification and validation of the model}: To determine how reliable the data generated by the model is, a verification and validation process will be conducted. In conjunction with the analysis of the dashboards extracted from the study of Ivankovic et al. \cite{Ivankovic2021}, the expertise acquired in the analysis and production of different dashboard prototypes and the dashboards design system at Infovis Public Health Research Project. Commencing with the guidelines delineated within the W3C-PROV standard and recent advancements in provenance research, this evaluation should encompasses an appraisal of the credibility of the proposed metadata constructs. Furthermore, should scrutinizes the model's potential for generalization across diverse contexts and its feasibility for replication, ensuring the reliability and reproducibility. The prototypes will be created for assessment by recognized experts in the field, aiming to gather their professional evaluation of the model. 


\item \textbf{Design an user interface for visualizing the dashboard provenance}: Analyzing provenance metadata is not a common task for the general public, particularly dashboard users, and requires literacy in provenance and provenance visualization \cite{Yazici2018}. Users may become overwhelmed in trying to understand and explore the meaning of data provenance due to its large volume and complexity. Data provenance visualization should be seamlessly integrated into the dashboard to enable satisfactory understanding. The use of infographics and storytelling approaches has been little explored in the literature of representing provenance, but it can be an appropriate solution for helping dashboard users to visualize provenance.  

\item \textbf{Assessment of the user's experience in navigating the dashboard's provenance  }: The assessment of the user's experience with the visualization of the dashboard provenance involves assessing how well the provenance information is presented and utilized within the dashboard interface, with the goal of understanding how users will perceive and interact with the provenance features and whether they will enhance their understanding and trust in the displayed information due to provenance. The evaluation should include user interviews, surveys, and user experience tests. The key aspects to consider during the evaluation are: user's understanding of the provenance information and its relevance to the dashboard content, the usability and accessibility of the provenance features within the dashboard interface, the impact on decision making, user overall satisfaction with the presentation of provenance information, and detect user preferences, suggestions for improvements, and whether they find the provenance features valuable and beneficial. This feedback can guide improvements and enhancements to enhance the user experience and ensure effective utilization of provenance in dashboards.
\end{itemize}

\section{Conclusion} 

This paper highlights the importance of incorporating provenance into dashboards to enhance the decision-making process. Provenance provides essential information about the dashboard's origin and the processes involved in its design, development, maintenance, and sustainability. It also identifies the various actors, such as institutions, departments, and teams, involved in the strategic and operational activities related to the dashboard's life-cycle. 

In contrast to traditional approaches that focus on representing the data in its final processed stage, provenance-driven dashboards enable users to see a more nuanced perspective on the curatorial history of the data and visualizations presented in the dashboard. By including provenance in dashboards, users can make more informed decisions based on a comprehensive understanding of the dashboard's context and reliability.

We have presented a model that aims to identify, capture, and organize the entities, agents, and activities, along with their attributes, properties, and relationships, to create a comprehensive model of dashboard provenance.

We recognize that will be necessary a lot of effort to create a comprehensive and thorough model of dashboard provenance. This is an ongoing endeavor and further progress is needed in the conclusion of the data provenance model to enable the design of interfaces for displaying the dashboard provenance in a standardized way. 

The model also requires thorough evaluation regarding its comprehensiveness and validity. This entails assessing the soundness of each suggested metadata construct, examining whether the model can be applied universally, and verifying if the model can be replicated to yield consistent outcomes.

We hope this paper will spark further development of dashboards with better provenance information that can positively impact the decision-making process.

\acknowledgments{%
	This study was financed in part by the Coordenação de Aperfeiçoamento de Pessoal de Nível Superior - Brasil (CAPES) - Finance Code 001%
}
\bibliographystyle{abbrv-doi-hyperref}

\bibliography{template}

%
%

\end{document}